\documentclass[letterpaper,english,aps,prl,twocolumn,floatfix,showpacs,amsfonts,amssymb]{revtex4}

\usepackage[T1]{fontenc}
\usepackage[latin1]{inputenc}
\usepackage{graphics}
\usepackage{amssymb}
\usepackage{overpic}

\newcommand{\beq}{\begin{equation}}
\newcommand{\eeq}{\end{equation}}
\newcommand{\bea}{\begin{eqnarray}}
\newcommand{\eea}{\end{eqnarray}}

\newcommand{\rmd}{{\rm d}}
\newcommand{\rmi}{{\rm i}}

\newcommand{\ucusi}{UCu$_{2}$Si$_{2}$}
\newcommand{\ucuge}{UCu$_{2}$Ge$_{2}$}
\newcommand{\ucusige}{UCu$_{2}$Si$_{2-x}$Ge$_{x}$}

\begin{document}

\title{Interplay between disorder, quantum and thermal fluctuations in 
ferromagnetic alloys: The case of UCu$_{2}$Si$_{2-x}$Ge$_{x}$}

\author{M.~B.~Silva~Neto}

\email{sneto@bu.edu}

\affiliation
{Department of Physics, Boston University, Boston, MA 02215}

\author{A.~H.~Castro~Neto}

\affiliation
{Department of Physics, Boston University, Boston, MA 02215}

\author{D.~Mixson}

\affiliation
{Department of Physics, University of Florida, Gainesville, FL 32611}

\author{J.~S.~Kim}

\affiliation
{Department of Physics, University of Florida, Gainesville, FL 32611}

\author{G.~R.~Stewart}

\affiliation
{Department of Physics, University of Florida, Gainesville, FL 32611}

\date{\today}

\begin{abstract}

We consider, theoretically and experimentally, the effects of 
structural disorder, quantum and thermal fluctuations in the 
magnetic and transport properties of certain ferromagnetic alloys.
We study the particular case of {\ucusige}. The low temperature 
resistivity, $\rho(T,x)$, exhibits Fermi liquid (FL) behavior as 
a function of temperature $T$ for all values of $x$, which can be 
interpreted as a result of the magnetic scattering of the conduction 
electrons from the localized U spins. The residual resistivity, 
$\rho(0,x)$, follows the behavior of a disordered binary alloy. 
The observed non-monotonic dependence of the Curie temperature, 
$T_c(x)$, with $x$ can be explained within a model of localized 
spins interacting with an electronic bath whose transport 
properties cross-over from ballistic to diffusive regimes. Our 
results clearly show that the Curie temperature of certain alloys 
can be enhanced due to the interplay between quantum and thermal 
fluctuations with disorder. 
\end{abstract}


\maketitle

Ternary intermetallic alloys of the type MT$_{2}$X$_{2}$, where 
M is an actinide or rare earth element, T is a transition metal 
and X is Si or Ge, have been a subject of intense 
experimental and also theoretical interest due to their interesting 
magnetic and transport properties \cite{Chelmicki}. In these systems, 
M is usually a magnetic atom with a partially filled $f$ shell, 
like U or Ce, and the conduction takes place in the $d$-bands of 
the transition metal such as Cu. These materials crystallize in 
the ThCr$_{2}$Si$_{2}$ tetragonal structure, with the $I4/mmm$ 
space group, and exhibit a rich variety of ground states 
\cite{Review}. For example, while {\ucusi} is a collinear 
ferromagnet with a high Curie temperature ($\sim 100$ K) 
\cite{Chelmicki}, URu$_{2}$Si$_{2}$ behaves like an Ising 
antiferromagnet \cite{Ian-Mydosh}, with strong crystalline 
fields \cite{CF-URu2Si2}, a low N\'eel temperature, and an 
anomalously small low temperature effective moment. Moreover, 
while the former is an ordinary metal at low temperatures, 
the latter is a superconductor. More generally, it has been 
noticed that both the magnetic and transport properties of 
such systems can be {\it classified} with respect to the 
smallest distance between two M atoms, a phenomenological 
parameter, $d_{h}$, called the {\it Hill} distance 
\cite{Hill,DeLong}. However, a theoretical explanation for 
the wealth of behavior observed in these materials is still 
lacking.

In recent years the interest in controlling the Curie 
temperature of diluted magnetic semiconductors (DMS) 
such as Ga$_{1-x}$Mn$_x$As has renewed the interest
on the nature of ferromagnetism in disordered alloys 
\cite{DMS}. However, DMS have many complications 
associated with the fact that the magnetic atom is also 
a donor/acceptor and its location is random in the 
lattice, leading to strong magnetic as well as structural 
disorder. Therefore, theoretical approaches to DMS are 
bound to be rather complex. In this work we follow a 
different route. We would like to separate the different 
physical mechanisms that control ferromagnetism in alloys. 
In {\ucusige} the magnetic sublattice is not directly 
affected by disorder which occurs only on the deep $p$ 
orbitals of Si/Ge. This structural disorder, however, 
affects the conduction band because of the $p-d$ 
hybridization. Furthermore, because Si and Ge are 
isovalent there is no affect on the carrier density. 
Therefore, this class of ferromagnetic materials allows 
the detailed study of the effect of structural disorder 
on the ferromagnetic properties without introducing extra 
complications. As we shall see, even this simplified 
problem has already unexpected behavior due to the 
interplay between structural disorder and quantum/thermal 
fluctuations \cite{Marcello-Antonio}. In particular we 
show that the unusual non-monotonic behavior of 
$T_{c}(x)$ as a function of $x$ can be understood as 
a cross-over between ballistic and diffusive electronic 
behavior. Therefore, one can show that by controlling the 
amount of disorder in the sample one can control directly 
the Curie temperature of the material. The repercussion 
of these results on DMS is obvious.  

The {\ucusige} alloys produced for this study were 
stoichiometrically weighed and melted together in a 
zirconium-gettered purified argon atmosphere. The 
polycrystalline samples were wrapped in Tantalum foil 
and annealed in sealed quartz glass tubes to insure
good crystalline order. The tetragonal structure of the 
polycrystalline samples was checked by X-ray powder 
diffraction at room temperature. We found that 
annealing at 875 $^{\circ}$C for $2$ weeks gave the 
best results (minimal weight loss and sharpest magnetic 
transitions). The samples were cut into rectangular 
bars ($0.8 \times 0.8 \times 8$ mm$^{3}$) for resistance 
measurements. Pt wires were spot welded to these samples. 
Electrical measurements used standard 4-wire dc techniques. 
The relative accuracy of the resistance measurements 
was approximately 4\%. Measurements of susceptibility 
were performed using a SQUID susceptometer in a field 
$B_{0}=1000$G. 

The two electrons on the $f$ level of U$^{4+}$, which is
assumed to be the configuration of the U atoms in the 
compound, are likely to be found in the $^{3}$H$_{4}$ Hund's 
multiplet configuration. The corresponding effective moment 
is $\mu_{eff}= 3.58 \mu_B$ in agreement with the 
measured moment at high temperatures in {\ucusi} \cite{Chelmicki}. 
Magnetic susceptibility, $\chi(T)$, measurements also suggest 
a small crystal field splitting at low temperatures which 
will be neglected in what follows. In {\ucuge}, on the other 
hand, the measured effective moment is $\mu_{eff}=2.40\mu_{B}$ 
\cite{Chelmicki}, which might be due to either crystal field 
effects or magnetic moment compensation (it may also be 
related to the antiferromagnetic transition observed at 
$T_{N}\sim 40$K). 

\begin{table}
\caption{Lattice parameters $a$ and $c$ of the
tetragonal structure ThCr$_{2}$Si$_{2}$ measured 
by X-ray powder diffraction; effective moment 
$\mu_{exp}$ determined out of $\chi(T)$ for $T>200$ K 
\cite{Chelmicki}; Fermi energy $E_{F}$ estimated 
by band structure calculations and measured from 
the bottom of the Cu $d$ band; ferromagnetic 
transition temperature $T_{c}$ determined from 
susceptibility measurements in a field $B_{0}=1000$G.}
\label{table-1}
\begin{center}
\begin{tabular}{l|c|c|c|c|c|c} \hline \hline
       & $a$ [{\AA}] & $c$ [{\AA}] & $\mu_{eff}^{J}
       [\mu_{B}]$ & $\mu_{exp} [\mu_{B}]$ & $E_{F}$ [eV] & $T_{c}$ [K] \\ \hline \hline
UCu$_{2}$Si$_{2}$ & $3.981$ & $9.939$ & $3.58$ & $3.58$ & $5.95$ & $102.5$ \\ \hline
UCu$_{2}$Ge$_{2}$ & $4.063$ & $10.229$ & $3.58$ & $2.40$ & $5.61$ & $108.55$  \\ \hline
\end{tabular}
\end{center}
\end{table}

\begin{figure}[htb]
\includegraphics[scale=0.28,angle=-90]{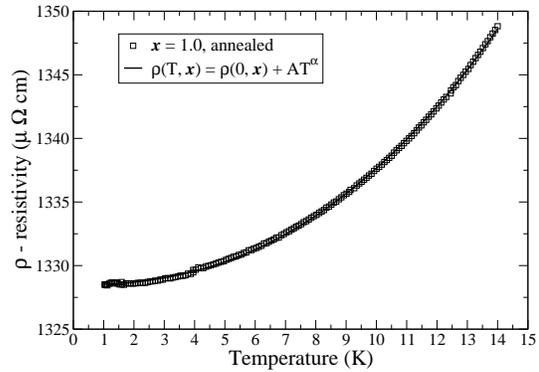}
\caption{Resistivity of {\ucusige} as a function of 
temperature for $x=1.0$. For the fit we used $\alpha=2.33$. 
Similar behavior for the resistivity is observed in all 
the other annealed samples where $\alpha$ did not 
significantly deviate from the FL value $\alpha=2$.}
\label{Fig-1}
\end{figure}

\begin{figure}[htb]
\includegraphics[scale=0.28,angle=-90]{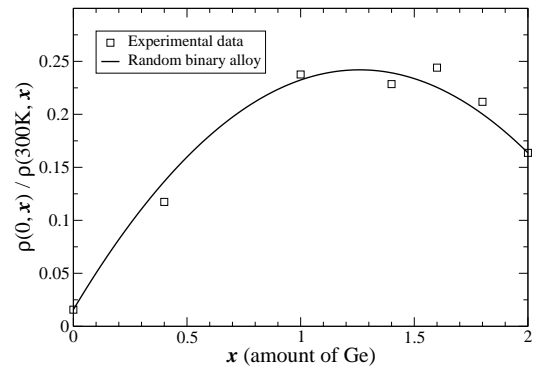}
\caption{Experimental data for the residual resistivity
of {\ucusige} normalized by its value at $300$ K (squares) 
and the fit corresponding to the binary alloy (solid line), 
according to the text.}
\label{Fig-2}
\end{figure}

The resistivity data suggests that the main scattering
mechanism at low temperatures is of magnetic origin, 
with the conduction electrons being scattered by magnetic
excitations \cite{Doniach} and has the FL form:  
$\rho(T,x)=\rho(0,x)+AT^{\alpha}$, with $\alpha\approx 2$ 
(see Fig. \ref{Fig-1}). $\rho(0,x)$ for a random binary 
alloy is expected to have the functional form:  
$\rho(0,x) = \rho_{Si}(2-x)/2 + \rho_{Ge} x/2 +
\bar{\rho} x(2-x)$, where $\rho_{Si}$ ($\rho_{Ge}$) is the
intrinsic resistivity of the pure compound at $x=0$ ($x=2$),
and the last term is the contribution from Nordheim's
rule \cite{Ziman}. However, we have found that the experimental value
of $\rho(0,x)$ does not follow this functional dependence with $x$.
The reason for this discrepancy is the existence of 
micro-cracks in the sample due to internal stresses generated
by doping. In order to eliminate their effect from the data we normalize
$\rho(0,x)$ by the high temperature resistivity, $\rho(300 K, x)$. 
In Fig.\ref{Fig-2} we show $\rho(0,x)/\rho(300K,x)$ as a function of $x$ 
together with the theoretical result with the above $x$ dependence. 
The experimental data, plotted in this way, are very well described 
by the random alloy expression. 

%
%
%

\begin{figure}[htb]
\includegraphics[scale=0.28,angle=-90]{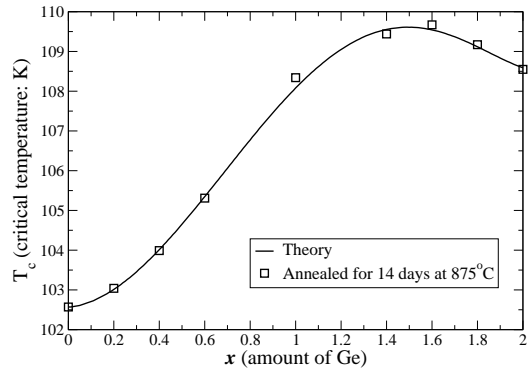}
\caption{Curie temperature, $T_{c}$, as a function of 
the amount of Ge, $x$. Experimental results: squares;
theoretical result from Eq. \ref{Gap-Equation}: solid line.}
\label{Fig-3}
\end{figure}

$T_{c}(x)$ was obtained from susceptibility measurements and a 
non-monotonic behavior as a function of $x$ was observed 
(see Fig.\ref{Fig-3}). A monotonic increase in $T_c$ is 
naively expected having in mind that the magnetic interaction 
in this system is RKKY and there is a shift of the Fermi energy
due to the change in the volume of the unit cell with $x$
(see below). However, the unexpected non-monotonicity in 
$T_c(x)$ was predicted recently by two of us (MBSN and AHCN) 
due to the interplay between fluctuations and structural 
disorder in alloys \cite{Marcello-Antonio}. The origin of 
such behavior can be tracked down to a change on the 
transport properties from ballistic (clean) to diffusive 
(dirty), depending on $x$. 

%
%
%

The physical properties of the two stoichiometric samples, {\ucusi}
and {\ucuge}, are shown in Table \ref{table-1}. The magnetic exchange 
interaction between U spins is ascribed to an isotropic RKKY interaction. 
Such interaction has already been shown to capture correctly the physics of 
these two compounds as well as of other isostructural systems such as 
URh$_{2}$Si$_{2}$ and UPd$_{2}$Si$_{2}$ \cite{Chelmicki}. Moreover, 
band structure calculations \cite{Denlinger} show a parabolic dispersion 
for the Cu $d$-band, at least along the direction of the tetragonal axis 
in the Brillouin zone. Assuming that alloying 
only introduces scattering centers for the conduction electrons
we propose the following Hamiltonian for these systems:
\bea
{\cal H}&=&
\sum_{{\bf k},\sigma}\varepsilon_{d}({\bf k})
d^{\dag}_{{\bf k},\sigma}d_{{\bf k},\sigma}+
\sum_{{\bf k},\sigma}V_{{\bf k},{\bf k}^{\prime}}d^{\dag}_{{\bf k},\sigma}
d_{{\bf k}^{\prime},\sigma^{\prime}}\nonumber\\
&+&\sum_{{\bf k},{\bf k}^{\prime}\sigma,\sigma^{\prime}}
{\cal J}({\bf k}){\bf J}_{-{\bf k}}\cdot
d^{\dag}_{{\bf k}^{\prime},\sigma}\vec{\sigma}_{\sigma\sigma^{\prime}}
d_{{\bf k}^{\prime}+{\bf k},\sigma^{\prime}}\nonumber\\
&+&U_{c}\sum_{{\bf q}}\rho_{{\bf q}}\rho_{-{\bf q}},
\label{Hamiltonian}
\eea
where ${\cal J}({\bf r}_{i}-{\bf r}_{j})=\sum_{{\bf k}}{\cal J}({\bf k})
e^{\rmi{\bf k}({\bf r}_{i}-{\bf r}_{j})}$ is an effective
exchange coupling due to the hybridization between the
$f$ states and conduction electrons, $d^{\dag}_{i,\sigma}$ 
($d_{i,\sigma}$) is the conduction electron creation (annihilation) 
operator for an electron with spin $\sigma=\uparrow,\downarrow$ 
on the $i^{th}$ site, 
${\bf J}_i = \sum_{{\bf k}} {\bf J}_{{\bf k}} e^{i {\bf k} \cdot {\bf r}_i}$ 
is the U magnetic moment on the $i^{th}$ site, 
$\vec{\sigma}_{\sigma\sigma^{\prime}}$ are 
the Pauli matrices, $\varepsilon_{d}({\bf k})$ is dispersion of the 
conduction band, $V_{{\bf k},{\bf k}^{\prime}}$ is a random 
impurity scattering potential, and $U_{c}$ is the Coulomb interaction 
between the conduction electrons ($\rho_{{\bf q}} = \sum_{{\bf k},\sigma}
c^{\dag}_{{\bf k}+{\bf q},\sigma} c_{{\bf k},\sigma}$ is the density
operator). The introduction of electron-electron
interactions in the conduction band is important because of the 
enhancement of the magnetic interactions due to the magnetic 
polarizability of the conduction band \cite{nick}. Furthermore,
although (\ref{Hamiltonian}) also describes the Kondo effect 
between the U spin and conduction band, the Kondo effect does not
play any role in the ferromagnetic phase \cite{larkin}.  
The above Hamiltonian (\ref{Hamiltonian}) can be derived from a 
more general three band Hamiltonian that includes not only the 
conduction electron $d$-band and $f$ states but also the $p$-band 
of Si/Ge with a random distribution of onsite energies 
\cite{Next-BU-Fl}.

As a first approach, we can perform a simple calculation in order 
to estimate $T_{c}(x)$ within a Weiss mean field theory of the 
Hamiltonian (\ref{Hamiltonian}) with $U_{c}=V=0$. We find (we use 
units where $\hbar=k_{B}=1$): 
\beq
T^{MF}_{c}= \frac{J(J+1)}{8}\frac{{\cal J}^{2}}{E_F}
\frac{N_{e}N_{m}}{N_{A}^{2}},
\label{Tc-mean-field}
\eeq
where $N_{e},N_{m}$, and $N_{A}$ are, respectively, the number of 
conduction electrons, magnetic ions, and total number of atoms, 
per unit cell, and $E_{F}$ is the Fermi energy. According to 
(\ref{Tc-mean-field}) the ratio between the Curie temperatures for 
the two stoichiometric samples, $T_c(x=2)/T_c(x=0)$, depends only 
on the ratio ${\cal J}^{2}/E_{F}$. From band structure calculations 
we estimate $E_{F}(x=0)=5.95$ eV and $E_{F}(x=2)=5.61$ eV, both 
measured with respect to the bottom of the Cu $d$ band. If we make 
the further assumption that ${\cal J}$ is not changed with $x$ we 
conclude that $T_c(x=2)/T_c(x=0)= E_F(x=0)/E_F(x=2)$ is fairly 
satisfied with the values of Table \ref{table-1}. This simple 
formulation of the problem, however, is not capable of explaining 
the non-monotonic behavior of $T_{c}(x)$ as shown in Fig. \ref{Fig-3}.

%
%
%

In a previous theoretical work \cite{Marcello-Antonio} it was 
shown that a continuum SU(N) formulation of the effective 
Heisenberg Hamiltonian, obtained after integration of the 
conduction electrons in (\ref{Hamiltonian}), that takes into 
account both the spin fluctuations and dissipation introduced 
by the electronic bath, might be a reasonable starting point 
to explain Fig. \ref{Fig-3}. In that formulation, the Curie 
temperature $T_{c}$ can be determined from the equation 
\cite{Marcello-Antonio}:
\beq
\frac{1}{g_{0}}=\int \frac{\rmd^{3}{\bf k}}{(2\pi)^{3}}
\int_{-\Omega^{0}_{k}}^{\Omega^{0}_{k}}\frac{\rmd\omega}{\pi}
\frac{k{\;}\gamma_{0}(kl){\;}\omega{\;}n_{B}(\omega/T_{c})}
{(k(k^{2}-c_{0}\omega))^{2}+(\gamma_{0}(kl){\;}\omega)^{2}}
\label{Gap-Equation}
\eeq
where $n_{B}(z)=(e^{z}-1)^{-1}$ is the Bose-Einstein distribution
function, $\Omega^{0}_{k}=(2k_{F})^{2}k/\gamma_{0}(kl)$ is an
energy cutoff, $g_{0}=Na/2{\cal J}_{H}S^{2}$ is the coupling constant, 
$c_{0}=1/{\cal J}_{H}Sa^{2}$ is the topological constant, and we have 
defined ${\cal J}_{H}={\cal J}^{2}(k_{F}a)^{3}/
4\pi^{2}E_{F}(1-a^{3}N(0)U_{c})$ 
which plays the role of the effective exchange. Furthermore,
$N(0)=m^{*}k_{F}/\pi^{2}$ is the density of states at the Fermi 
energy $E_{F}=k_{F}^{2}/2m^{*}$, $k_{F}=(3\pi^{2}n)^{1/3}$ is the 
Fermi wave vector, and $n=N_{e}/a^{3}$ is the electronic density
per unit cell with lattice spacing $a$. Finally, the dissipation 
coeficient is given by \cite{fulde-luther}:
\beq
\gamma_{0}(kl)=\frac{8k_{F}^{2}}{v_{F}(1-a^{3}N(0)U_{c})}
\frac{\mbox{arctan}(kl)}{1-(kl)^{-1}\mbox{arctan}(kl)},
\label{Dissipation-Rate}
\eeq
where $l$ is the electronic mean free path, and 
$v_{F}=k_{F}/m^{*}$ is the Fermi velocity.

Eq. (\ref{Gap-Equation}) relates $T_{c}$ to a number of 
material properties and reduces to Eq. (\ref{Tc-mean-field}) 
in the limit of $\gamma_{0}\rightarrow 0$ and 
$J\rightarrow\infty$ \cite{Next-BU-Fl}. 
Each Cu$^{2+}$ contributes one electron to the conduction 
band and we can set $N_{e}=8$. The replacement of Si by 
Ge alters the size of the unit cell, thus we use Vegard's 
law to write $a=a_{Si}(2-x)/2+a_{Ge}x/2$, where $a_{Si}=5.4$ 
{\AA} and $a_{Ge}=5.527$ {\AA}, were chosen in order to 
reproduce the volume of the unit cell (see Table \ref{table-1}). 
Notice that the product $k_{F}a$ does not depend on $x$. We 
also fix the values of $m^{*}_{Si}$ and $m^{*}_{Ge}$ so that 
the calculated $E_{F}$ matches the values obtained from band 
structure calculations, and we use $J=4$, $N=2$ and $U_{c}=0.45$ 
eV, consistent with estimates from Thomas-Fermi theory
\cite{Next-BU-Fl}. 

The electronic mean free path can be obtained from 
the Drude relation: $\rho=m^{*}v_{F}/ne^{2}l$ ($e$ is the charge 
of the electron). From the residual resistivity (see Fig. \ref{Fig-2}) 
we obtain the dependence of $l$ on $x$ 
\cite{Next-BU-Fl}. Typical values for the product $k_{F}l$ 
range from $k_{F}l\sim 10$ for $x=0$, where the
system is in the ballistic regime (note that $\gamma_{0}(k_{F}l\gg 1)\propto$
constant), to $k_{F}l\sim 0.74$ for $x=1.3$, in the 
diffusive regime ($\gamma_{0}(k_{F}l\ll 1)\propto 
1/k_{F}l$). In particular, $k_{F}l\simeq 0.8$ for $x=1.6$ where 
$T_{c}$ is the highest \cite{Next-BU-Fl}. 
Notice also that the maximum in $T_c(x)$ in Fig. 
\ref{Fig-3} is very close to the position where the resistivity 
(mean free path) is largest (shortest) in Fig. \ref{Fig-2} 
indicating that $T_c$ is mostly controlled by electron scattering. 
Thus, as discussed in \cite{Marcello-Antonio}, since $T_{c}$ of 
a dissipative spin fluctuation system is larger in the diffusive 
relative to the ballistic regime, we can understand the non-monotonic 
behavior of $T_{c}(x)$ in Fig. \ref{Fig-3} as a direct consequence 
of the behavior of the residual resistivity in Fig. \ref{Fig-2}. An 
enhancement of the Curie temperature in the disordered single band 
Hubbard model has also been reported recently in Ref. 
\cite{Volhardt}.

Finally, because of the changes in the unit cell volume, we expect 
${\cal J}(x)$ to be monotonically decreasing with $x$ \cite{Next-BU-Fl}. 
For the stoichiometric samples, {\ucusi} and {\ucuge}, ${\cal J}$ 
can be estimated from a hybridization model for the magnetic-ordering 
behavior in U intermetallic compounds like the one considered here 
\cite{Mydosh}. According to Ref. \cite{Mydosh} the hybridization is 
governed by the $f$-$d$ hybridization, $V_{fd}$, the strength of 
which is a function of the ionic radius of U$^{4+}$ and Cu$^{2+}$, 
their angular momentum quantum numbers, $l$ and $m$, and their 
relative distance in the unit cell. The exchange parameter, 
${\cal J}$, is then calculated from $V_{fd}$ by means of a 
Schrieffer-Wolf transformation \cite{Mydosh}. Typical values 
found for ${\cal J}$ are between $8$ meV ($x=0$) and $3.5$ 
meV ($x=2$). For $0<x<2$, however, the situation is much more 
complex due to the effects of disorder in the hybridization 
matrix elements. Nevertheless, we find that ${\cal J}$ is a 
monotonically decreasing function of $x$ so that the all the 
non-monotonicity in $T_{c}(x)$ can be uniquely attributed to 
the disorder introduced by the conduction electron scattering 
potential. The final theoretical value of $T_c(x)$, from Eq. 
\ref{Gap-Equation}, is shown as the continuous line on Fig.
\ref{Fig-3}. 

In conclusion, we have studied, theoretically and experimentally,
the effect of structural disorder in the magnetic and transport
properties of ferromagnetic alloys of the form {\ucusige}. We
have shown that the interplay between disorder and magnetism leads 
to an unexpected non-monotonic behavior of the Curie temperature 
that cannot be explained by naive mean field theories. We have
shown that in order to describe the ferromagnetic ordering in these 
systems one needs to take into account the dissipation introduced 
by the coupling of the local spins to the electronic heat bath. 
Dissipation is controlled by the electron mean-free path and 
therefore is sensitive to the amount of disorder and the 
electron-electron interactions. The same physical processes can 
be important in the physics of DMS. However, the introduction of 
magnetic disorder is very important and has to be carefully considered. 

MBSN and AHCN are deeply indebted to E.~Novais for his criticism
and support. The authors have also benefited from illuminating  
discussions with K.~Bedell, P.~Dalmas~de~R\'eotier, S.~Das~Sarma, 
L.~DeLong, P.~F.~Farinas, and D.~Volhardt. M.~B.~Silva~Neto 
acknowledges CNPq  (Brazil) for financial support and Boston 
University for the hospitality. Work at Florida performed under 
the auspices of a DOE grant DE-FG05-86ER45268.

\end{document}